\documentclass{llncs}

\usepackage[utf8]{inputenc}

\usepackage{tikz}
\usetikzlibrary{decorations.markings}
\usetikzlibrary{arrows}
\usepackage{amssymb}
\usepackage{amsfonts,amsmath,latexsym,stmaryrd}
\usepackage{listings}
\usepackage{cite}
\usepackage{alltt}
\usepackage{rotating}
\usepackage{pdflscape}
\usepackage{booktabs}
\usetikzlibrary{shapes,decorations}
\usepackage{verbatim} 
\usepackage{graphicx}
\usepackage{fancyvrb}
\usepackage{algorithm}
\usepackage{algpseudocode}
\usepackage{algorithmicx}
\usepackage{float}
\usepackage{url}

\usepackage{todonotes}

\algnewcommand\algorithmicinput{\textbf{Input:}}
\algnewcommand\INPUT{\item[\algorithmicinput]}

\algnewcommand\algorithmicoutput{\textbf{Output:}}
\algnewcommand\OUTPUT{\item[\algorithmicoutput]}

\newtheorem{PSM}[theorem]{Premise Selection Method}

\newcommand{\C}[1]{\mbox{\lstinline`#1`}}

\definecolor{dkblue}{rgb}{0,0.1,0.5} 
\definecolor{lightblue}{rgb}{0,0.5,0.5} 
\definecolor{dkgreen}{rgb}{0,0.4,0} 
\definecolor{dk2green}{rgb}{0.4,0,0} 
\definecolor{dkviolet}{rgb}{0.6,0,0.8}
\definecolor{shadethmcolor}{rgb}{0.9, 0.9,1}

\begin{document}

\title{Proof Mining with Dependent Types\thanks{The work was supported by EPSRC grants EP/J014222/1 and EP/K031864/1.}}

\author{Ekaterina Komendantskaya \and  J\'onathan Heras}
\authorrunning{J. Heras and E. Komendantskaya}

\institute{Mathematical and Computer Sciences, Heriot-Watt University, UK\\
\email{ek19@hw.ac.uk}
\and
Mathematics and Computer Science,
University of La Rioja, 
Rioja, Spain\\
\email{joheras@gmail.com}
}

\maketitle

\begin{abstract}
Several approaches exist to data-mining big corpora of formal proofs. Some of these approaches
are based on statistical machine learning, and some -- on theory exploration.
However, most are developed for either untyped or simply-typed theorem provers.  
In this paper, we present a method that combines statistical data mining and theory exploration in order to analyse and
automate proofs in  
 dependently typed language of Coq.
\end{abstract}

\section{Introduction}\label{sec:introduction}


Interactive Theorem Provers (ITP) are functional programming languages for writing and verifying proofs in type theory. Some of them, like Coq or Agda, feature 
 dependent and  higher-order inductive types.
The ITP community has developed several methods to improve automation, e.g. special tactic languages (\emph{Ltac} in Coq~\cite{Coq})
or special libraries (e.g. SSReflect~\cite{SSReflect}).
Some provers are hybrid between automated theorem proving and ITP, e.g.  ACL2~\cite{KMM00-1}. 
The Isabelle/HOL community pioneered methods of direct interfacing of interactive provers with third-party automated provers~\cite{BlanchetteKPU16}, which can also work for a subset of Coq~\cite{CzajkaK16}.


The large volume of proof data coupled with growing popularity of machine-learning tools 
inspired an alternative approach to improving automation in ITP.
 Machine-learning can be used to learn proof strategies  from existing proof corpora. This approach has been tested in different ITPs: in Isabelle/HOL via its connection with Sledgehammer~\cite{BlanchetteKPU16} or via the library Flyspeck~\cite{KaliszykU15}, in ACL2~\cite{HerasKJM13} and in 
Mizar~\cite{UrbanRS13}. For Coq, however, only partial solutions have been suggested~\cite{KHG13,HK14,GWR14,GWR15}. 

Several challenges  arise when data mining proofs in Coq:
\begin{itemize}
\item[C1.] Unlike e.g. ACL2 or Mizar, Coq's types play a role in proof construction and the proposed  machine-learning methods should account for that.
\item[C2.] Unlike e.g. Isabelle/HOL, Coq has \emph{dependent types}, thus separation between proof term and type level syntax is impossible. Any machine-learning method for Coq should reflect the dependency between terms and types.
\item[C3.] Coq additionally has a tactic language \emph{Ltac} introduced to facilitate interactive proof construction. This feature being popular with users, it also needs to be captured in our machine-learning tools.
\end{itemize}

Challenge C3 was tackled in~\cite{KHG13,HK14,GWR14,GWR15}, but we are not aware of any existing solutions to challenges C1-2.  
It was shown in ML4PG~\cite{KHG13,HK14} that clustering algorithms can analyse similarity between \emph{Ltac} proofs and group them into clusters. Generally, this knowledge can be used either to aid the programmer in proof analysis~\cite{HK14}, or as part of heuristics for construction of new similar proofs~\cite{HerasKJM13}. 
This paper enhances ML4PG's clustering methods with the analysis of proof-term structure, which takes into consideration mutual dependencies between terms and types, as well as the recursive nature of the proofs.
The novel method thus addresses challenges C1-2. 
To complete the picture, we also propose a new premiss selection algorithm for Coq that generates new proof tactics in the \emph{Ltac} language based upon the clustering output, thus addressing C3 in a new way. 

Together, the proposed proof clustering and the premiss selection methods offer a novel approach to proof automation with dependent types: \emph{``Data mine proof terms, generate proof tactics"}. This allows to access and analyse the maximum of information at the data mining stage, and then reduce the search space by working with a simpler tactic language at the proof generation stage.

\section{Bird's Eye View of the Approach and Leading Example}\label{sec:example}


\textbf{1. Data mining Coq proof terms.}
In Coq, to prove that a theorem $A$ follows from a theory (or current proof context) $\Gamma$, we would need to \emph{construct}
an inhabitant $p$ of type $A$, which is a proof $p$ of $A$ in context $\Gamma$;  denoted by $\Gamma \vdash p: A$. Sometimes, $p$ is also called the \emph{proof term} for $A$.
A type checking problem ($\Gamma \vdash p:A?$) asks to verify whether $p$ is indeed a proof for $A$, a type inference problem ($\Gamma \vdash p:?$) asks to verify whether the alleged proof is a proof at all; and a type inhabitation problem ($\Gamma \vdash ? : A$ ) asks to verify whether $A$ is provable. 
Type inhabitation problem is undecidable in Coq, and the special tactic language \emph{Ltac} is used to aid the programmer in constructing the proof term $p$.

We illustrate the interplay of Coq's syntax and the  \emph{Ltac} tactic language syntax in the following example.
Consider the following proof of associativity of $\mathtt{append}$ that uses \emph{Ltac} tactics:

\begin{lstlisting}
Theorem app_assoc: forall l m n:list A, l ++ m ++ n = (l++m)++ n.
Proof.  
intros l m n; induction l; simpl; trivial.
rewrite IHl; trivial.
Qed.
\end{lstlisting}

What we see as a theorem statement above is in fact the type of that proof, and the tactics merely help us to construct the proof term that inhabits this type, which we can inspect by using \lstinline?Print app_assoc.? command:

\begin{lstlisting}
app_assoc = fun l m n : list A =>
list_ind (fun l0 : list A => l0 ++ m ++ n = (l0 ++ m) ++ n) eq_refl
(fun (a : A) (l0 : list A) (IHl : l0 ++ m ++ n = (l0 ++ m) ++ n) =>
eq_ind_r (fun l1 : list A => a :: l1 = a :: (l0 ++ m) ++ n) eq_refl IHl) l
     : forall l m n : list A, l ++ m ++ n = (l ++ m) ++ n
\end{lstlisting}
Because Coq is a dependently typed language,  proof terms and types may share the signature. 
Context $\Gamma$ above is given by the \texttt{List} library defining, among other things, the two list constructors  and operation of $\mathtt{append}$.

\emph{The first methodological decision we make here is to data mine proof terms, rather than \emph{Ltac} tactics.}


\textbf{2. Capturing the structural properties of proofs by using term trees.}
\emph{The second decision we make when analysing the Coq syntax is to view both proof term and type associated with a Coq's object as a tree.} This makes structural properties of proofs apparent.

To use an example with a smaller tree, consider the Coq Lemma \lstinline?forall (n : nat) (H : even n), odd (n + 1)?.
 Its term tree is depicted in Figure~\ref{fig:termtree}.

\begin{figure}[t]
\centering
\begin{tikzpicture}[level 1/.style={sibling distance=40mm},
 level 2/.style={sibling distance=70mm},
 level 3/.style={sibling distance=30mm},
 level 4/.style={sibling distance=30mm},scale=.8,font=\footnotesize]
   
   \node (root) {\lstinline?forall?} [level distance=10mm]
             child { node {\lstinline?n : nat?}}
             child { node {\lstinline?H : even n?}}
             child { node {\lstinline?odd : nat -> Prop?} 
                      child { node {\lstinline?+ : nat -> nat -> nat?}
                           child { node {\lstinline?n : nat?}}
                           child { node {\lstinline?1 : nat?}}
                           }
                   }
            
    ;
 \end{tikzpicture}
\caption{\scriptsize{\emph{Coq term tree for the pCIC term \texttt{forall (n : nat) (H : even n), odd (n + 1).}}}}\label{fig:termtree}
\end{figure} 

\textbf{3. Clustering  term trees.} \emph{We next decide on a suitable machine learning approach to analyse Coq proofs, and settle on unsupervised learning.}    
\emph{Clustering algorithms}~\cite{Bishop} have become a standard tool for finding patterns in large data sets. We use the term \emph{data object} to refer to an individual element within a data set. In our case, the data set is given by Coq libraries, and the data objects are given by term trees obtained from Coq terms and types. Theorem \lstinline?app_assoc? is one object in the List library.

Clustering algorithms divide data objects into similar groups (clusters), based on (numeric) vector representation of the objects' properties.
The process of extracting such  properties  from data is called 
 \emph{feature extraction}~\cite{Bishop}. \emph{Features} are parameters chosen to represent all objects in the data set. \emph{Feature values} are usually given by numbers that instantiate the features for every given object. 
If $n$ features are chosen to represent all objects, the features form \emph{feature vectors} of length $n$; and clustering is conducted in an $n$-dimensional space.

\textbf{3.1. Converting tree structures  to feature matrices.}
We now need to decide how to convert each proof tree into a feature vector. 
A variety of methods exists to represent trees as matrices, for instance using
adjacency or incidence matrices. The adjacency matrix method shares the following common properties with various previous methods of  feature
extraction for proofs~\cite{lpar-urban,K13}: different 
library symbols are represented by distinct features, and the 
feature values are binary. 
For 
large libraries and growing syntax, such feature vectors grow very large (up to $10^6$ in some experiments).

We propose an alternative method that can characterise a large (potentially unlimited) number of Coq terms by a finite number of features and a (potentially unlimited) number of feature values.
The features that determine the rows and columns of the matrices  are given by the term tree depth and the level index of nodes. 
In addition, given a Coq term, its features must differentiate its term and type components. 
As a result, in our encoding
each tree node is reflected by three features
that represent the term component and the type component of the given node, as well as the level index of its parent node in the term tree,
cf. Table~\ref{ml4pgtermtable} (in which $td$ refers to ``tree depth"). 

\begin{table}[t]
\caption{\footnotesize{\emph{Term tree matrix for \texttt{forall (n : nat) (H : even n), odd (n+1)}.}}\label{ml4pgtermtable}}{
\centering
{\scriptsize 
\begin{tabular}{| p{0.5cm} || p{4.2cm} | p{3cm} | p{4cm} |}
\hline
 & level index 0 & level index 1 & level index 2 \\
 \hline
td0 & ($[\mathtt{forall}]_{Gallina}$,-1,-1) & (0,0,0) & (0,0,0)\\
\hline
td1 & ($[\mathtt{n}]_{term}$,$[\mathtt{nat}]_{type}$,0) & ($[\mathtt{H}]_{term}$,$[\mathtt{even\ n}]_{type}$,0)& ($[\mathtt{odd}]_{term}$,$[\mathtt{nat\rightarrow Prop}]_{type}$,0) \\
\hline
td2 & ($[\mathtt{+}]_{term}$,$[\mathtt{nat \rightarrow nat \rightarrow nat}]_{type}$,2) & (0,0,0)& (0,0,0)\\
\hline
td3 & ($[\mathtt{n}]_{term}$,$[\mathtt{nat}]_{type}$,0) & ($[\mathtt{1}]_{term}$,$[\mathtt{nat}]_{type}$,0)  & (0,0,0)\\
\hline
\end{tabular}}}
\end{table}

\textbf{3.2. Populating feature matrices with feature values.}
Feature matrices give a skeleton for extracting proof properties. Next, it is important to have an intelligent algorithm to populate the feature matrices with rational numbers -- i.e. \emph{feature values}.
Consider the associativity of the operation of list append given in Theorem \lstinline?app_assoc?.  Associativity is a property common to many other operations. For example, addition on natural numbers is associative:

\begin{lstlisting}
Theorem plus_assoc: forall l m n: nat, l + (m + n) = (l + m) + n.
Proof.
intros l m n; induction l; simpl; trivial.
rewrite IHl; trivial.
Qed.
\end{lstlisting}
The corresponding proof term is as follows:
\begin{lstlisting}
plus_assoc = fun l m n : nat =>
nat_ind (fun l0 : nat => l0 + (m + n) = l0 + m + n) eq_refl
  (fun (l0 : nat) (IHl : l0 + (m + n) = l0 + m + n) =>
   eq_ind_r (fun n0 : nat => S n0 = S (l0 + m + n)) eq_refl IHl) l
     : forall l m n : nat, l + (m + n) = l + m + n
\end{lstlisting}

We would like our clustering tools to group theorems \lstinline?app_assoc? and \lstinline?plus_assoc? together. It is easy to see that the term tree structures of the theorems \lstinline?app_assoc? and \lstinline?plus_assoc? are similar, and so will be the structure of their feature matrices. However, the feature matrix cells are populated by feature values.
If values assigned to \lstinline?list? and \lstinline?nat? as well as to  \lstinline?++? and \lstinline?+? are very different, then these two theorems 
may not be grouped into the same cluster.  

To give an example of a bad feature value assignment, suppose we define functions $[.]_{term}$ and $[.]_{type}$ to blindly assign a new natural number to every new object defined in the library. Suppose further that we defined natural numbers at the start of the file of 1000 objects, and lists -- at the end of the file, 
then we may have an assignment $[$\lstinline?nat?$]_{type} = 0$, $[$\lstinline?+?$]_{term} =1$, $[$\lstinline?list?$]_{type} = 999$, $[$\lstinline?++?$]_{term} = 1000$. This assignment would suggest to treat functions \lstinline?+? and \lstinline?++? as very distant. 
If these values populate the feature matrices for our two theorems \lstinline?app_assoc? and \lstinline?plus_assoc?, the corresponding feature vectors will lie in distant regions of the $n$-dimensional plane. This may lead the clustering algorithm to group these theorems in different clusters.


 Seeing that  the definitions of \lstinline?list? and \lstinline?nat? are structurally similar, and so are \lstinline?++? and \lstinline?+?, we would rather characterise their similarity by close feature values, 
 irrespective of where in the files they occur. 
But then our algorithm needs to be intelligent enough 
to calculate how  similar or dissimilar these  terms are. For this, we need to cluster all objects in the chosen libraries, find that \lstinline?list? and \lstinline?nat? are clustered together, and  \lstinline?++? and \lstinline?+? are clustered together, and assign close values to objects from the same cluster.
   In this example, definitions of \lstinline?list?, \lstinline?nat?, \lstinline?++? and \lstinline?+?  do not rely on any other definitions (just Coq primitives), and we can extract their feature matrices and values directly (after manually assigning values to a small fixed number of Coq primitives). In the general case, it may take more than one iteration to reach the primitive symbols. 
	
We call this method of mutual clustering and feature extraction \emph{recurrent clustering}. It is extremely efficient in analysis of dependently-typed syntax, in which inductive definitions and recursion play a big role.
It also works uniformly to capture the entire Coq syntax: not just proofs of theorems and lemmas in \emph{Ltac} like in early versions of ML4PG, 
but also all the type and function definitions.  
   
	\textbf{4. Using clusters to improve proof automation.}
	Suppose now we have the output of the clustering algorithm, in which all Coq objects of interest are grouped into $n$ clusters. All objects in each cluster share some structural similarity, like \lstinline?app_assoc? and \lstinline?plus_assoc?.
	Suppose we introduce a new lemma $L$ for which we do not have a proof but which is clustered together with \lstinline?app_assoc? and \lstinline?plus_assoc?. We then can re-adjust the sequences of \emph{Ltac} tactics used in proofs of \lstinline?app_assoc? and \lstinline?plus_assoc? and prove $L$.

	Our running example so far was simple, and the tactics used in the proofs did not refer to other lemmas. The new recurrent clustering method is most helpful when proofs actually call some other auxiliary lemmas. Simple  recombination of proof tactics does not work well in such cases: our new  lemma $L$ clustered with other finished proofs in cluster $C$ may require similar, but not exactly the same auxiliary lemmas. 
	
Take for example  lemma \lstinline?maxnACA? that states the inner commutativity of the maximum of two natural numbers in the SSReflect library \lstinline?ssrnat?:

\begin{lstlisting}
Lemma maxnACA : interchange maxn maxn. 
\end{lstlisting}


ML4PG clusters  \lstinline?maxnACA? together with already proven lemmas $\{$\lstinline?addnACA?, \lstinline?minnACA?, \lstinline?mulnACA?$\}$ --- these
 three lemmas state the inner commutativity of addition, multiplication and the minimum of two naturals, respectively. 
We will try to construct the proof for  \lstinline?maxnACA? by analogy with the proofs of $\{$\lstinline?addnACA?, \lstinline?minnACA?, \lstinline?mulnACA?$\}$.
 Consider the proof of the lemma \lstinline?addnACA? in that cluster: it is proven using the sequence of tactics \lstinline?by move=> m n p q; rewrite -!addnA (addnCA n).? That is, it mainly relies on auxiliary lemmas \lstinline?addnA? and \lstinline?addnCA?.
The proof of \lstinline?addnACA? fails to apply to \lstinline?maxnACA? directly. In particular, 
the auxiliary lemmas
	\lstinline?addnA? and \lstinline?addnCA? 
	 do not  apply in the proof of \lstinline?maxnACA?.  But similar lemmas would apply! And here the clustering method we have just described comes to the rescue.
	If we cluster all auxiliary lemmas used in the proofs of $\{$\lstinline?addnACA?, \lstinline?minnACA?, \lstinline?mulnACA?$\}$ with the
 rest of the lemmas of the \texttt{ssrnat} library we find the cluster $\{$\lstinline?minnA?, \lstinline?mulnA?, \lstinline?maxnA?, \lstinline?addnA? $\}$  and the cluster $\{$\lstinline?minnAC?, \lstinline?mulnAC?, \lstinline?maxnAC?,  \lstinline?addnCA?$\}$.
These two clusters give us candidate auxiliary lemmas to try in our proof of \lstinline?maxnACA?, in places where \lstinline?addnA? and \lstinline?addnCA? were used in \lstinline?addnACA?.
As it turns out, the lemmas \lstinline?maxnA? and \lstinline?maxnAC? will successfully apply as auxiliary lemmas in \lstinline?maxnACA?, and the sequence of tactics \lstinline?by move=> m n p q; rewrite -!maxnA (maxnCA n)?  proves the lemma \lstinline?maxnACA?.
 


\textbf{Paper overview.} The rest of the paper is organised as follows. Section~\ref{sec:background} introduces some of the background concepts.
In Section~\ref{sec:R1}, we define the algorithm for automatically extracting features matrices from Coq terms and types. 
In  Section~\ref{sec:R2}, we define a second algorithm, which automatically computes feature values to populate the feature matrices. 
These two methods are new, and have never been presented before. 
In Section~\ref{sec:PSM}, we propose the new method of premiss selection and tactic generation for Coq based on clustering.
Finally, in Section~\ref{sec:RW} we survey related work and conclude the paper.

\section{Background}\label{sec:background}

The underlying formal language of Coq is known as the \emph{Predicative Calculus of (Co)Inductive Constructions} (pCIC)~\cite{Coq}.

\begin{definition}[pCIC term]\label{def:coqterms}

$-$ The sorts \lstinline?Set?, \lstinline?Prop?, \lstinline?Type(i)? (\lstinline?i?$\in \mathbb{N}$) are terms.

$-$ The global names of the environment are terms. 

$-$ Variables are terms.
 
$-$ If \lstinline?x? is a variable and \lstinline?T, U? are terms, then \lstinline?forall x:T,U? 
 is a term. If \lstinline?x? does not occur in \lstinline?U?, then \lstinline?forall x:T,U? will be written as \lstinline?T -> U?. A term of the 
 form \lstinline?forall x1:T1, forall x2:T2, ..., forall xn:Tn, U? will be written as \lstinline?forall (x1:T1) (x2:T2) ...(xn:Tn), U?. 
 
$-$ If \lstinline?x? is a variable and \lstinline?T, U? are terms, then \lstinline?fun x:T => U? 
 is a term. A term of the form \lstinline?fun x1:T1 => fun x2:T2 => ... => fun xn:Tn => U? will be written as \lstinline?fun (x1:T1) (x2:T2) ...(xn:Tn) => U?. 

 $-$ If \lstinline?T? and \lstinline?U? are terms, then \lstinline?(T U)? is a term -- we use an uncurried notation (\lstinline?(T U1 U2 ... Un)?) 
 for nested applications (\lstinline?(((T U1) U2) ... Un)?).
 
 $-$ If \lstinline?x? is a variable, and \lstinline?T, U? are terms, then (\lstinline?let x:=T in U?) is a term.

\end{definition}

The syntax of Coq~\cite{Coq} includes some terms that do not appear in Definition~\ref{def:coqterms}; e.g. given 
 a variable \lstinline?x?, and  terms \lstinline?T? and \lstinline?U?, \lstinline?fix name (x:T) := U? is a Coq term used to declare a recursive definition.
The notion of a term in Coq covers a very general syntactic
category in the Gallina specification language. 
However, for the purpose of concise exposition, 
we will restrict our notion of a term to Definition~\ref{def:coqterms},
giving the full treatment of the whole Coq syntax in the actual ML4PG implementation.

\textbf{Clustering.} A detailed study of performance  of different well-known clustering algorithms in proof-mining in ML4PG can be found in~\cite{KHG13}.  
In this paper we use the $k$-means clustering algorithm throughout
as it gave the best evaluation results in~\cite{KHG13} and is rather fast (we use an implementation in Weka~\cite{Weka}).

The chosen clustering algorithm relies on a user-supplied parameter $k$ that identifies the number 
of clusters to form. 
Throughout this paper, we will use the following heuristic method~\cite{KHG13,HerasKJM13} to determine $k$:
$$k=\lfloor\frac{\text{objects to cluster}}{10-g}\rfloor,$$
where $g$ is a user-supplied granularity value from $1$ to $5$. Lower granularity ($1-2$) indicates the general preference for a smaller  number of clusters of larger size, and higher granularity ($3-5$) suggests 
a larger number of clusters of smaller size. In this setting, $g$ is supplied by the user, via the ML4PG interface, as an indicator of a general intent. 

\section{Feature Extraction, Stage-1: Extracting Feature Matrices from pCIC Terms}\label{sec:R1}

We  first introduce a suitable tree representation of pCIC terms. We refer to Section~\ref{sec:example} for examples supporting the  definitions below.

\begin{definition}[pCIC term tree]\label{def:ml4pgtermtree}
Given a pCIC term \lstinline?C?, we define its associated \emph{pCIC term tree} as follows:

$-$ If \lstinline?C? is one of the sorts \lstinline?Set?, \lstinline?Prop? or \lstinline?Type(i)?, then the  pCIC term tree of 
 \lstinline?C? consists of one single node, labelled respectively by \lstinline?Set:Type(0)?, \lstinline?Prop:Type(0)? or \lstinline?Type(i):Type(i+1)?.
 
$-$ If \lstinline?C? is a name or a variable, then 
 the pCIC term tree of \lstinline?C? consists of one single node, labelled by the name or the variable itself together with its type.
 
$-$ If \lstinline?C? is a pCIC term of the form \lstinline?forall (x1:T1) (x2:T2) ...(xn:Tn), U? (analogously for \lstinline?fun (x1:T1) (x2:T2) ...(xn:Tn) => U?); 
then, the term tree of \lstinline?C? is the tree with the root node labelled by \lstinline?forall? (respectively \lstinline?fun?) 
and its immediate subtrees given by the trees representing \lstinline?x1:T1?, \lstinline?x2:T2?, \lstinline?xn:Tn? and \lstinline?U?.

$-$ If \lstinline?C? is a pCIC term of the form \lstinline?let x:=T in U?, then the pCIC tree of \lstinline?C? 
 is the tree with the root node labelled by \lstinline?let?, having three subtrees given by the trees corresponding to \lstinline?x?, \lstinline?T? and \lstinline?U?.

$-$ If \lstinline?C? is a pCIC term of the form  \lstinline?T -> U?, then  the pCIC term tree of \lstinline?C?  is represented by the tree with the root node labelled by 
\lstinline?->?, and its immediate subtrees given by the trees representing  \lstinline?T? and \lstinline?U?.

$-$ If \lstinline?C? is a pCIC term of the form  \lstinline?(T U1 ... Un)?, then we have two cases.
If \lstinline?T? is a name, the pCIC term tree of \lstinline?C?  is represented by 
the tree with the root node labelled by \lstinline?T? together with its type, and its immediate subtrees given by the trees
representing \lstinline?U1?,\ldots, \lstinline?Un?. If \lstinline?T? is not a  name, the pCIC term tree of \lstinline?C?
is the tree with the root node labelled by \lstinline?@?, and its immediate subtrees given by the trees 
representing \lstinline?T?, \lstinline?U1,...,Un?.

\end{definition}

Note that pCIC term trees extracted from any given Coq files consist of two kinds of nodes: \emph{Gallina} and \emph{term-type} nodes. Gallina is a specification language of Coq, it contains keywords and special tokens such as \lstinline?forall?, \lstinline?fun?, \lstinline?let? or \lstinline?->? (from now on, we will call them Gallina tokens). 
 The term-type nodes are given by expressions of the form \lstinline?t1:t2? where \lstinline?t1? is a sort, a variable or 
a name, and  \lstinline?t2? is the type of \lstinline?t1?.

We now convert pCIC term trees into feature matrices:

\begin{definition}[pCIC Term tree depth level and level index]\label{def:termtreelevel}
Given  a pCIC term tree $T$, the \emph{depth} of the node $t$ in $T$, denoted by \emph{depth(t)}, is defined as follows:

$-$ $depth(t) = 0$, if $t$ is a root node;

$-$ $depth(t) = n+1$, where $n$ is the depth of the parent node of $t$.

The \emph{$n$th level} of $T$ is the ordered sequence of nodes of depth $n$.  As is standard,  the order of the sequence is 
given by visiting the nodes of depth $n$ from left to right. 
 We will say that the size of this sequence is the \emph{width} of the tree at depth $n$. The width of $T$ is given by the largest width of its levels.
The \emph{level index} of a node with depth $n$ is the position of the node in the $n$th level of $T$.
We denote by $T(i,j)$ the node of $T$ with depth $i$ and index level $j$.
\end{definition}

We use the notation $M[\mathbb{Q}]_{n\times m}$ to denote the set of matrices of size  $n\times m$ with rational coefficients.

\begin{definition}[pCIC term tree feature matrix]\label{df:matrix}
Given a pCIC term \lstinline?t?, its corresponding pCIC term tree $T_{\mathtt{t}}$ with the depth $n$ and the width $m$, and three injective functions 
$[.]_{term}: pCIC~terms \rightarrow \mathbb{Q}^+$, $[.]_{type}: pCIC~terms \rightarrow \mathbb{Q}^+$
and $[.]_{Gallina}: Gallina~tokens \rightarrow \mathbb{Q}^-$, \emph{the feature extraction function} 
$[.]_M=\langle[.]_{term}, [.]_{type},[.]_{Gallina}\rangle : pCIC~terms \rightarrow M[\mathbb{Q}]_{n \times 3m}$
builds the  \emph{term tree matrix of $\mathtt{t}$}, $[\mathtt{t}]_M$,
where the $(i,j)$-th entry of $[\mathtt{t}]_M$ captures information from the node $T_{\mathtt{t}}(i,j)$ as follows:

$-$ if $T_{\mathtt{t}}(i,j)$ is a Gallina node $g$, then the $(i,j)$th entry of $[\mathtt{t}]_M$ is a triple $([g]_{Gallina},-1,p)$ 
where $p$ is the level index of the parent of $g$. 

$-$ if $T_{\mathtt{t}}(i,j)$ is a $term:type$ node \lstinline?t1:t2?, then the $(i,j)$th entry of $[\mathtt{t}]_M$ is a triple $([t1]_{term},[t2]_{type},p)$
where $p$ is the level index of the parent of the node.
\end{definition}

One example of a term tree feature matrix for the term tree of \lstinline?forall (n : nat) (H : even n), odd (n + 1)? is given in Table~\ref{ml4pgtermtable}. Since the depth of the tree is $4$, and its width is $3$, it takes the matrix of the size $4 \times 9$ to represent that tree.
Generally, if the largest pCIC term tree in
	the given Coq library has the size $n \times m$, we take  feature matrices of size $n \times 3 m$ for all feature extraction purposes.  The resulting feature vectors have an average density ratio of 60\%. It has much smaller feature vector size, and much higher density than  in sparse approaches~\cite{lpar-urban,K13}. It helps to obtain more accurate clustering results.

In Definition~\ref{df:matrix}, we specify the functions $[.]_{Gallina}, [.]_{term}$ and $[.]_{type}$ just by their signature. 
The function $[.]_{Gallina}$ is a predefined function.
The number of Gallina tokens is fixed and cannot be 
expanded by the Coq user. Therefore, we know in advance all the Gallina tokens that can appear in a development, and we can
assign a concrete value to each of them. The function $[.]_{Gallina}: Gallina~tokens  \rightarrow \mathbb{Q}^-$ is an injective 
function  defined to assign close values to similar Gallina tokens 
and more distant numbers to unrelated tokens. 

The functions $[.]_{term}$ and $[.]_{type}$ are dynamically re-defined for every 
library and every given proof stage, as the next section will explain.

\section{Feature Extraction Stage-2: Assigning Feature Values via Recurrent Clustering}\label{sec:R2}

When defining the functions
 $[.]_{term}$ and $[.]_{type}$, 
we must ensure that these functions are sensitive to the structure of terms, assigning close values 
to similar terms, and more distant values to dissimilar terms.

Starting with the primitives, $[.]_{term}$ and $[.]_{type}$ will always assign fixed values to the pre-defined 
sorts in Coq (cf.  item (1) in Definition~\ref{def:funterm}). Next, suppose $t$ is the $n$th object of the given Coq library.  For variables $x_1, \ldots, x_n$ occurring in a term $t$, $[x_i]_{term} = i$ and $[x_j]_{type} = j$, using a method resembling de Brujn indices (cf. item (2)).
Item (3)  defines $[.]_{term}$ and $[.]_{type}$ for the recursive calls.
The most interesting (recurrent) case occurs when $[.]_{term}$ and $[.]_{type}$ need to be defined for subterms of $t$ that are defined elsewhere in the library.
In such cases, we use output of recurrent clustering of the first $n-1$ object of the library.
When the first $n-1$ Coq objects are clustered, each cluster is automatically assigned a unique integer number. Clustering algorithms additionally assign 
a \emph{proximity value} (ranging from $0$ to $1$) to every object in a cluster to indicate the proximity of the cluster centroid, or in other terms,  the certainty of the given example belonging 
to the cluster. The definitions of $[.]_{term}$ and $[.]_{type}$ below  use the cluster number and the proximity value to compute feature values:

\begin{definition}\label{def:funterm}
Given a Coq library, its $n$th object given by a pCIC term  \lstinline?t?, the corresponding pCIC term tree $T_{\mathtt{t}}$, and a node $s = T_{\mathtt{t}}(k,l)$,  the functions $[.]_{term}$ and $[.]_{type}$ are defined respectively for the term component \lstinline?t1?
and the type component \lstinline?t2? of $s$  as follows:

 \begin{enumerate}

\item (Base case) If \lstinline?t1? (or \lstinline?t2?) is the $i$th element of the set $\{$
\lstinline?Set, Prop, Type(0),? \lstinline?Type(1), Type(2) ? $\ldots \}$
then  its value $[\mathtt{t1}]_{term}$ (or $[\mathtt{t2}]_{type}$) is given by $100+\sum_{j=1}^i\frac{1}{10\times 2^{j-1}}$.

\item (Base case) If \lstinline?t1? (or \lstinline?t2?) is the $i$th distinct variable in \lstinline?t?, then $[\mathtt{t1}]_{term}$ (or $[\mathtt{t2}]_{type}$) is assigned the value $i$.  

\item (Base case) If  \lstinline?t1?$=$\lstinline?t?, i.e. \lstinline?t1? is a recursive call (or 
\lstinline?t2?$=$\lstinline?t?, i.e. \lstinline?t? is an inductive type definition), we assign a designated constant to  \lstinline?t1? (or \lstinline?t2?, respectively).  


\item (Recurrent case) If \lstinline?t1? or \lstinline?t2?  is an $m$th object of the given Coq libraries, where $m < n$,
then  cluster the first $n-1$ objects of that library and take into account the clustering output as follows.

If \lstinline?t1? (or \lstinline?t2?) belongs to a cluster $C_j$ with associated proximity value $p$,  then $[\mathtt{t1}]_{term}$ (or 
$[\mathtt{t2}]_{type}$, respectively) is assigned the value $200+2\times j + p$.

\item (Recurrent case) If \lstinline?t1:t2? is a local assumption or hypothesis, then cluster this object against  the first $n-1$ objects of the given libraries and take into account the clustering output as follows. If \lstinline?t1? (or \lstinline?t2?) belongs to a cluster $C_j$ with associated proximity value $p$,  then $[\mathtt{t1}]_{term}$ (or $[\mathtt{t2}]_{type}$, respectively) is assigned the value $200+2\times j + p$. 

\end{enumerate}
 
\end{definition}

In the formula for sorts (item 1), the component $\sum_{j=1}^i\frac{1}{10\times 2^{j-1}}$ produces small fractions to reflect the similarity of all sorts, and 
$100$ is added in order to  distinguish sorts from variables and names.
The formula $200+2\times j + p$ used elsewhere assigns $[\mathtt{t}]$ a value within $[200+2\times j,200+2\times j+1]$ depending on the
 proximity value of \lstinline?t?  in cluster $j$. Thus, elements of the same cluster have closer values compared to the values 
assigned to elements of other clusters. For example, using the three clusters shown below, $[\mathtt{eqn}]_{term} = 200+2\times1+0.5 = 202.5$, where $0.5$ is the proximity value of \lstinline?eqn?
in Cluster 1. By contrast, $[\mathtt{drop}]_{term} = 200+2\times2+0.7 = 204.7$, where $0.7$ is a proximity value of \lstinline?drop? in Cluster 2. 
 
\textbf{The Algorithm at Work: Examples}

We finish this section with some examples of clusters discovered in the basic infrastructure of the SSReflect 
library~\cite{SSReflect}.
We include here 3 of the 91 clusters discovered by our method automatically after processing 457 objects (across 12 standard files), within 5--10 seconds.

\begin{itemize}
 \item Cluster 1:
{\small \begin{lstlisting}
 Fixpoint eqn (m n : nat) :=
   match m, n with 
   | 0, 0 => true | m'.+1, n'.+1 => eqn m' n' 
   | _, _ => false end.
 Fixpoint eqseq (s1 s2 : seq T)  :=
   match s1, s2 with 
   | [::], [::] => true | x1 :: s1', x2 :: s2' => (x1 == x2) && eqseq s1' s2' 
   | _, _ => false end.         
\end{lstlisting}}
    
\item  Cluster 2:
{\footnotesize 
\begin{lstlisting}
 Fixpoint drop n s := match s, n with | _ :: s', n'.+1 => drop n' s' | _, _ => s end.
 Fixpoint take n s := match s, n with | x :: s', n'.+1 => x :: take n' s' | _, _ => [::] end.
\end{lstlisting}}

\item Cluster 3:
{\small 
\begin{lstlisting}
 Definition flatten := foldr cat (Nil T).
 Definition sumn := foldr addn 0.
\end{lstlisting}} 
\end{itemize}


The first cluster contains the definitions of equality for natural numbers and lists --- showing that 
the clustering method can spot structural similarities across different libraries. The second cluster discovers similarity between \lstinline?take? (takes the first $n$ elements of 
a list) and \lstinline?drop? (drops the first $n$ elements of a list). 
The last pattern is less trivial of the three, as it depends on 
other definitions, like \lstinline?foldr?, \lstinline?cat? (concatenation of lists) and \lstinline?addn? (sum
of natural numbers). 
Recurrent term clustering handles such dependencies well: it assigns close values to \lstinline?cat? and \lstinline?addn?,  since they have been discovered to belong to the same cluster. 
Note the precision of the recurrent clustering. Among $457$ terms it considered, $15$ used \lstinline?foldr?, however,  Cluster 3 contains only $2$ definitions, excluding e.g. 
\lstinline?Definition allpairs s t:=foldr (fun x => cat (map (f x) t)) [::] s? ; \lstinline?Definition divisors n:=foldr add_divisors [:: 1] (prime_decomp n)?  or \lstinline?Definition Poly:=foldr cons_poly 0.?  This precision is due to the recurrent clustering process with its deep analysis of the term and type structure, including analysis of any  auxiliary functions. This is how it discovers that functions \lstinline?add_divisors? or \lstinline?cons_poly? are structurally different from  auxiliary functions \lstinline?cat? and \lstinline?addn?, and hence definitions \lstinline?allpairs?, \lstinline?divisors? and \lstinline?Poly? are not included in Cluster 3. 

This deep analysis of term structure via recurrent clustering improves accuracy and will play a role in the next section.

\section{Applications of Recurrent Proof Clustering. A Premiss Selection Method}\label{sec:PSM}

Several premiss selection methods have been proposed for Isabelle, HOL, as well as several other provers:
~\cite{KaliszykU15,UrbanRS13}. 
Relying on Coq's tactic language, and the clustering results of the previous section, we can formulate the problem of \textbf{premiss selection for Coq} as follows:
\emph{Given a cluster $C$ of pCIC objects from a Coq library $L$ and an arbitrary theorem/lemma $T$ in this cluster, 
can we recombine sequences of proof tactics used to prove other theorems/lemmas in $C$ in such a way as to  obtain a proof for $T$? In particular, if the proof of $T$ requires the use of auxiliary Lemmas from $L$, can we use the outputs of the recurrent clustering to make valid suggestions of auxiliary lemmas? }

The  algorithm below answers these questions in the positive. To answer the second question, we suggest to automatically examine all
 auxiliary lemmas $A_1, \ldots , A_m$ used in proofs that belong to the  cluster $C$, and look up the clusters to which $A_1, \ldots , A_m$ belong in order to suggest auxiliary lemmas for $T$. This is the essence of the clustering-based premiss selection method we propose here (see especially the item 4(b) and Algorithm~\ref{PSM-a}   below). As Section~\ref{sec:example} has illustrated, often a combination of auxiliary lemmas is required in order to complete a given proof, and the below algorithm caters for such cases.

\begin{PSM}\label{pem:ml4pg}
Given the statement of a theorem $T$ and a set of lemmas $L$, find a proof for $T$ as follows:
\begin{enumerate}
 \item Using the recurrent clustering and $T\cup L$ as dataset, obtain the cluster $C$ that contains the theorem $T$ 
 (possibly alongside other lemmas $L'_1, L'_2,  \ldots L'_n$).
 \item Obtain the sequence of tactics $\{T_1^j,\ldots,T_{n_i}^j\}_{j}$ that are used to prove each lemma $L'_j$ in $C$.
 \item Try to prove $T$ using $T_1^j,\ldots,T_{n_j}^j$, for each $j \in \{1, \ldots n\}$.
 \item If no such sequence of tactics proves $T$, then infer new arguments for each tactic $T_k^j$ as follows:
    \begin{enumerate}
     \item If the argument of $T_k^j$ is an internal hypothesis from the context of a proof, try all the internal hypothesis from the context of the current proof. 
     \item If the argument of $T_k^j$ is an external lemma $E$, use the recurrent clustering and $L$ as dataset to determine which  lemmas are in a cluster with $E$ and try all those lemmas in turn as arguments of $T_k^j$.
     \item[***] This can be naturally extended to tactics with several arguments, as Algorithm~\ref{PSM-a}  shows.
  \end{enumerate}    
\end{enumerate}
\end{PSM}

The heart of the above procedure is the process of Lemma selection (item 4(b)), which we state more formally as Algorithm~\ref{pem:ml4pg}.

\begin{algorithm}                      
\caption{Premiss selection method based on clustering (item 4(b) of Method~\ref{pem:ml4pg})}          
\label{PSM-a}                           
\begin{algorithmic}[1]                    
\Require $\mathcal{S}$ -- a set of lemmas and theorems contained in the given Coq libraries $\mathcal{L}_1, \ldots, \mathcal{L}_s$.
\Require $T(L_1, \ldots, L_k)$ -- an Ltac tactic $T$ with arguments containing $L_1, \ldots, L_k \in \mathcal{S}$.  
\Require $\mathcal{S}/C_n$ partition of the set $\mathcal{S}$ into $n$ clusters $C_1, \ldots, C_n$.
\State Denote by $L^j_{C_i}$ the $j$th lemma of cluster $C_i$.
\State The algorithm returns a list of suggested lemmas in place of each $L_l \in  (L_1, \ldots, L_k)$.
 \For{$l = 1, \ldots , k $}
    \For{$ i = 1, \ldots , n $}
		\If{$L_l \in C_i$}
            \For{$j = 1, \ldots \|C_i\|$}
                  \State $L_k$ := $[L_k :: L^j_{C_i}]$
            \EndFor
         \EndIf				     		
	 \EndFor
     \State \Return ($L_k$)
  \EndFor
\end{algorithmic}
\end{algorithm}

\textbf{Evaluation}

Using five Coq libraries of varied sizes, we perform an empirical evaluation of the premiss selection method, and thereby testify the accuracy of the proposed recurrent clustering. Our test data consists of the basic infrastructure of the SSReflect library~\cite{SSReflect}, the formalisation of Java-like bytecode presented in~\cite{HK14}, the formalisation of persistent homology~\cite{HCMS12}, the Paths library of the HoTT development~\cite{Hott},
 and the formalisation of the Nash Equilibrium~\cite{nash}. 

\begin{table}[H]
\caption{\footnotesize{\emph{Percentage of proofs discovered by analogy with similar proofs in the given library, for bigger ($g=1$) and smaller ($g=5$) clusters.}}\label{tab:reproven}}{
\centering
\begin{tabular}{l p{1.6cm} p{1.7cm} p{1.7cm} p{1.7cm} p{1.7cm}}
 \hline
 Library & Language & Granularity $g=1$ & Granularity $g=3$ & Granularity $g=5$ & Library size (No of Theorems) \\
 \hline
 SSReflect library  & SSReflect & $36\%$ & $35\%$ & $28\%$ & 1389\\
  JVM & SSReflect  & $56\%$ & $58\%$ & $65\%$ & 49\\
 Persistent Homology & SSReflect  & $0\%$ & $10\%$ & $12\%$ & 306\\
 Paths (HoTT) & Coq & $92\%$ & $91\%$ & $94\%$ &80\\
 Nash Equilibrium & Coq & $40\%$ & $37\%$ & $36\%$ & 145\\
   \hline
\end{tabular}}
\end{table}

The results of our experiments are given in Table~\ref{tab:reproven}. The success rate of the premiss selection method 
depends on how similar the proofs of theorems in a given library are. 
Additionally, and unlike all other premiss selection methods~\cite{KaliszykU15,UrbanRS13}, we now have to factor in the fact that Coq's proofs as given by the tactics in \emph{Ltac} language may require a sophisticated combination of tactics for which our premiss selection method does not cater. Indeed, some proofs may not call auxiliary lemmas at all and incorporate all the reasoning within one proof.  
This explains the high success rate in the \emph{Paths (HoTT)}
library, where most of the lemmas are proven in the same style, and using auxiliary lemmas in the same well-organised way,  and the low rate in the \emph{Persistent Homology} 
library, where just a few lemmas have similar proofs with auxiliary lemmas. The granularity value does not have a big impact in the performance of 
our experiments, and almost the same number of lemmas is proven with different granularity values. In some cases, like in the Nash equilibrium
library, a small granularity value generates bigger clusters that increase the exploration space allowing to prove more lemmas. 
However, reducing the granularity value can also have a negative impact; for instance, in the JVM library the number of clusters
is reduced and this leads to a reduction in the number of the proven theorems.

\section{Conclusions and Related Work}\label{sec:RW}

The presented hybrid method of proof mining
combines statistical machine learning and premiss selection, Coq's proof terms and \emph{Ltac} tactics.  
It is specifically designed to cater for a dependently typed language such as that of Coq.
The \emph{recurrent clustering}   analyses tree structures of proof terms and types, and groups all Coq objects (definitions, lemmas, theorems, ...) into $n$ sets, depending on their dependencies and structural similarity with the given libraries. 
Previous versions of ML4PG could only analyse tactics, rather than proof terms or definitions. 

The output of the clustering algorithm can be used to directly explore the similarity of lemmas and definitions, and ML4PG includes a graphical interface to do that. In this paper we presented a novel method of premiss selection that allows to use the output of clustering algorithms for analogy-based premiss selection. 
We use the \emph{Ltac} language to automatically generate candidate tactics from clusters. 
  \emph{Ltac}'s language gives a much simpler syntax to explore compared with a possible alternative -- a direct generation of Coq term trees.  

Evaluation of the method shows that it bears promise, especially in the libraries where the proofs are given in a uniform style, and auxiliary lemmas are used consistently. Capturing the role of auxiliary lemmas in proofs is known to be a hard problem in proof mining, and recurrent clustering gives a good solution to this problem.
Other existing methods, like~\cite{GWR14,GWR15} address limitations of our premiss selection method, and suggest more sophisticated algorithms for tactic recombination. Integration of these methods with ML4PG is under way.

The recurrent clustering method is embedded into ML4PG: \url{http://www.macs.hw.ac.uk/~ek19/ML4PG/}.
The  integration of the Premiss Selection method into ML4PG or directly into Coq's \emph{Ltac} 
is still future work.

\textbf{Related work} \\
\emph{Statistical machine learning for automation of proofs: premiss selection.}
 The method of statistical proof-premise selection~\cite{KaliszykU15,UrbanRS13,lpar-urban,K13}
is applied in several theorem provers. 
 The method we presented in Section~\ref{sec:PSM} is an alternative method of premiss selection, due to its reliance on the novel algorithm of recurrent clustering. Also, this paper presents the first premiss selection method that takes into consideration the structure of proof terms in a  dependently typed language.

\emph{Automated solvers in Coq.}
In~\cite{CzajkaK16}, the standard \emph{``hammer"} approach was adopted, where a fragment of Coq syntax  was translated into FOL, and then the translated lemma and theorem statements (without proof terms) were proven using existing first-order solvers. By contrast, our approach puts emphasis on mining proof terms that inhabit the lemmas and theorems in Coq. 
Our method does not rely on any third party first-order solvers. Results of clustering of the Coq libraries are directly used to reconstruct Coq proofs, by analogy with other proofs in the library. Thus, the process of (often unsound and inevitably incomplete) translation~\cite{CzajkaK16} from the higher-order dependently typed language of Coq into FOL becomes redundant.

\emph{Theory exploration: models from tactic sequences.}
While statistical methods focus on extracting information from existing large libraries, \emph{symbolic methods} are instead concerned with automating the discovery of lemmas in new  theories~\cite{GWR14,GWR15,JDB11,HLW12,hipster}, relying on existing proof strategies, e.g. proof-planning and rippling \cite{Duncan02,BB05}. 
The method of tactic synthesis given in Section~\ref{sec:PSM} belongs to that group of methods.

The method developed in this paper differs from the cited papers in that it allows to reason deeper than the surface of the tactic language. I.e., statistical analysis of pCIC proof terms and types as presented in Section~\ref{sec:R1} is employed before  the tactics in the \emph{Ltac} language are generated.

\emph{Combination of Statistical and Symbolic Machine Learning Methods in theorem proving.}
Statistical machine learning was used to support auxiliary lemma formulation by analogy in the setting of ACL2, see~\cite{HerasKJM13}.
In this paper, we extended that approach in several ways: we included types, proof terms and the tactic language -- all crucial building blocks of Coq proofs as opposed to first-order untyped proofs of ACL2. 

\bibliographystyle{plain} 
\bibliography{biblio}




\end{document}